\pdfoutput=1 
\documentclass[aps,prd,twocolumn,superscriptaddress,amsfont,graphicx,nofootinbib,preprintnumbers]{revtex4}

\usepackage{amsmath}
\usepackage{amssymb}
\usepackage{color}
\usepackage{xcolor}
\usepackage{graphicx}
\usepackage{hyperref}
\usepackage{textcomp}
\usepackage{xspace,slashed}
\usepackage[utf8]{inputenc}
\usepackage{subcaption}
\usepackage{multirow}
\usepackage{nccmath} 
\usepackage{slashed}
\usepackage{orcidlink}

\newcommand{\ep}{\epsilon}
\newcommand{\amp}{\mathcal{A}}
\newcommand{\IR}{\mathcal{I}}
\newcommand{\UV}{\mathcal{U}}

\newcommand{\ddk}{\mathcal{N}}
\newcommand{\T}{\textbf{T}}
\newcommand{\ci}{\mathrm{i}}
\newcommand{\gram}{\mathrm{G}}

\newcommand{\UZHaff}{Physik-Institut, Universit\"at Z\"urich, Winterthurerstrasse 190, 8057 Z\"urich, Switzerland}

\begin{document}

\title{
The integrand form of infrared singularities of two-loop QCD scattering amplitudes
}

\author{Piotr Bargie\l{}a~\orcidlink{0000-0002-3646-5892}}
\email{piotr.bargiela@physik.uzh.ch}
\affiliation{\UZHaff}

\preprint{ZU-TH 33/25}

\begin{abstract}
In this work, we express the singular part of a scattering amplitude in terms of Feynman integrals compatible with topologies appearing in the bare amplitude, and we choose a basis of locally finite Master Integrals.
In two-loop massless QCD, we find such a representation of the amplitude singularities using a systematic ansatz reconstruction of the integrand from a predicted integrated form.
As an example application, we write the finite part of an amplitude for the digluon production in quark annihilation for some helicity configurations as manifestly locally finite.
\end{abstract}

\maketitle

\tableofcontents

\section{Introduction}
\label{sec:intro}

The most fundamental description of a probability amplitude for a scattering process of particles in a Quantum Field Theory (QFT) is given by scattering amplitudes.
They can be expressed as a perturbative expansion in terms of multi-loop Feynman diagrams in the small coupling region.
Such bare loop amplitudes are usually singular when defined explicitly in four spacetime dimensions.
In phenomenological applications, the most common procedure which allows to avoid such singularities is dimensional regularization (dimReg) in $d=4-2\ep$ around $\ep \to 0$.
In gauge theories, these singularities are universal i.e. they can be predicted without computing the whole amplitude.
Importantly, physical quantities are free of singularities.
Indeed, according to the Kinoshita, Lee, and Nauenberg (KLN) theorem~\cite{Kinoshita:1962ur,Lee:1964is}, contributions from virtual and real quantum corrections in any scattering process cancel all singularities.
For this reason, it is convenient to subtract the whole singular part of the amplitude from a bare amplitude such that all the physical information is contained in the finite part of the amplitude.

In the following, we will focus on the massless Quantum Chromodynamics (QCD) corrections to two-loop order.
The ultraviolet (UV) singularities arise from coupling renormalization, while the infrared (IR) singularities can be extracted from correlators of Wilson lines in soft-collinear effective theory (SCET).
The finite part of an amplitude is obtained by acting with UV and IR operators, defined later in the text, on lower-loop corrections and then subtracting these terms from the bare amplitude.
Due to the fact that the UV and IR operators predict only the singular poles $\ep^{-n}$ of the amplitude, they can be defined freely up to a finite part $\mathcal{O}(\ep^0)$.
This nonuniqueness leads to the existence of multiple available UV renormalization and IR regularization schemes, e.g. Catani~\cite{Catani:1998bh} or Becher-Neubert~\cite{Becher:2009cu,Becher:2009qa} scheme.

The main goal of this work is to express the singular part of the amplitude in terms of Feynman integrals compatible with topologies appearing in the bare amplitude, and then to choose a basis of locally finite Master Integrals, thus aiming for a locally finite expression for the finite part of the amplitude.
To this end, we propose a new scheme, to which we will refer as \textit{pure}, defined such that the UV and IR operators have a form compatible with the Feynman integrals appearing in the bare amplitude.
At the moment, available schemes allow for this property only for the UV operators but not the IR.
Indeed, the integrand of the amplitude consists of propagators involving squared loop momenta, while SCET correlators only involve propagators linearized in loop momenta.
For this reason, the IR operators are provided in the literature in their integrated form as an expansion in powers of $\ep$.
Contrarily, we derive the pure IR operator at two loops in a form of a linear combination of Feynman integrands with rational function coefficients such that, upon integrating over the loop momenta, they reproduce the expected $\ep^{-n}$ IR poles.
In practice, our systematic ansatz reconstruction of the integrand expression from its integrated form requires a careful disentanglement of kinematics from color, as well as of rational from transcendental contributions to the IR poles.
It also relies on understanding the minimal basis~\cite{Chicherin:2020oor,Chicherin:2021dyp,Gehrmann:2024tds} property of Feynman integrals.

Formulating scattering amplitudes purely in four spacetime dimensions is of a major modern interest across different collaborations.
It is due to the fact that this would lead to reducing the redundancy in the current formalism by making it more explicitly physical.
At the same time, it would be more appropriate for stable and efficient numerical evaluation.
As such, it would have vast applications to the automation of amplitude computations beyond the well-understood one-loop order.
Locally finite Feynman integrals can be constructed  e.g. by inserting appropriate numerators, as in Refs~\cite{Arkani-Hamed:2010zjl,Gluza:2010ws,Arkani-Hamed:2010pyv,Gambuti:2023eqh,delaCruz:2024xsm,Ma:2025mog}.
This has been already applied in two-loop five-point amplitude calculations, e.g. in Refs~\cite{Badger:2016ozq,Hartanto:2019uvl}.
An algorithm for finding a basis of locally finite Master Integrals has been introduced in Refs~\cite{vonManteuffel:2014qoa,vonManteuffel:2015gxa}.
It has been successfully applied to two-loop four-point amplitude calculations, e.g. in Refs~\cite{Borowka:2016ypz,vonManteuffel:2017myy,Bonetti:2020hqh,Agarwal:2020dye,Chen:2020gae}.
For loop-induced two-loop amplitudes with colorless final states, the local integrand subtraction has been found in Refs~\cite{Anastasiou:2018rib,Anastasiou:2020sdt,Anastasiou:2022eym,Anastasiou:2024xvk}.
Numerically, the Loop Tree Duality approach allows for local subtraction of UV and IR singularities, see Refs~\cite{Catani:2008xa,Runkel:2019yrs,Capatti:2019ypt,Kermanschah:2024utt}.
In this work, we use the locally finite integrals introduced in Ref.~\cite{Gambuti:2023eqh} to form a Master Integral basis for the locally finite part of an example QCD amplitude with four colored particles.

\section{UV renormalization and IR regularization}
\label{sec:scheme}

In this section, we review the standard way of extracting from a bare scattering amplitude its finite part in a  generic scheme.

Consider an amplitude $\amp$ for some multi-particle scattering process in a given QFT.
For definiteness, consider a QFT with a single coupling $g_s$.
Nonetheless, the general argument presented in what follows should hold also for QFTs with multiple couplings.
In order to regulate all of the UV and IR singularities stemming from loop corrections to the amplitude $\amp$, we will use dimReg in $d=4-2\ep$ dimensions, with $\ep \to 0$.
In the region of a small value of the coupling, the amplitude $\amp$ can be perturbatively expanded as
\begin{equation}
    \amp 
    = g_{s,b}^{2q} ( \amp_{0,b} 
    + g_{s,b}^2 \mu_0^{2\ep} \amp_{1,b} 
    + \left( g_{s,b}^2 \mu_0^{2\ep} \right)^2 \amp_{2,b} 
    + \mathcal{O} \left( g_{s,b}^4 \right) ) \,,
	\label{eq:ampBare}
\end{equation}
where $b$ denotes a bare quantity, $2q$ is a process-dependent nonnegative integer which depends on the leading order correction, and $\mu_0$ is the dimReg scale.
In this work, we will not consider corrections higher than $L=2$ loops, where our counting is relative to the leading order.
The UV singularities of the amplitude $\amp$ can be treated with the coupling renormalization via
\begin{equation}
	\alpha_{s,b} 
	= \alpha_{s}^{(S)}( \UV_{0}^{(S)}
	- \frac{\alpha_{s}^{(S)}}{2\pi} \UV_{1}^{(S)}
	+ \frac{(\alpha_{s}^{(S)})^2}{(2\pi)^2} \UV_{2}^{(S)}
	+ \mathcal{O} ( ( \alpha_{s}^{(S)} )^3 )) \,,
	\label{eq:asRen}
\end{equation}
where $\alpha_{s} = \frac{g_{s}^2}{4\pi}$, and $S$ denotes the chosen renormalization scheme.
We are going to keep track of the scheme $S$-dependence, in order to be able to compare between the schemes later on.
We introduced generic \textit{UV operators} $\UV_{L}^{(S)}$ which depend on $\ep$ and color invariants, have poles up to $\ep^{-L\ep}$, and act trivially in the color space.
Upon substitution, the amplitude can be expressed as a renormalized perturbative series 
\begin{equation}
	\amp 
	= \frac{(\alpha_{s}^{(S)})^q}{(2\pi)^q} ( \amp_{0}^{(S)}
	+ \frac{\alpha_{s}^{(S)}}{2\pi} \amp_{1}^{(S)} 
	+ \frac{(\alpha_{s}^{(S)})^2}{(2\pi)^2} \amp_{2}^{(S)}
	+ \mathcal{O} ( ( \alpha_{s}^{(S)} )^3 ) ) \,.
	\label{eq:ampRen}
\end{equation}
The renormalized amplitudes $\amp_{L}^{(S)}$ can be related to the bare amplitudes $\amp_{L,b}^{(S)}$ as
\begin{equation}
	\begin{split}
		\frac{1}{c_0} \amp_{0}^{(S)} &= \amp_{0,b} \,, \\
		\frac{1}{c_0} \amp_{1}^{(S)} &= \frac{\ci \, \UV_0^{(S)} \, \mu^{2\ep}}{2 f} \amp_{1,b} 
		- \frac{q \, \UV_1^{(S)}}{\UV_0^{(S)}} \amp_{0,b} \,, \\
		\frac{1}{c_0} \amp_{2}^{(S)} &= - \frac{\ci \, (\UV_0^{(S)})^2 \, \mu^{4\ep}}{4 f^2} \amp_{2,b} 
		+ \frac{\ci \, (1+q) \, \UV_1^{(S)} \, \mu^{2\ep}}{2 f} \amp_{1,b} \\
		&- \frac{q \, (q-1) \, (\UV_1^{(S)})^2 + 2 q \, \UV_0^{(S)} \, \UV_2^{(S)}}{2(\UV_0^{(S)})^2} \amp_{0,b} \,,
	\end{split}
	\label{eq:ampRen2bare}
\end{equation}
where we define $c_0 = (8 \pi^2 \mu^2 \mu_0^{-2} \, \UV_0^{(0)})^q$, and the loop factor $f=\frac{\ci}{16\pi^2}$.
We define a finite part of the amplitude $\amp_{L,\text{fin}}^{(S)}$ by acting nontrivially in the color and kinematic space on renormalized amplitudes $\amp_{L}^{(S)}$ with \textit{IR operators} $\IR_L^{(S)}$ which cancel all the remaining IR singularities that start at order $\ep^{-2L}$, i.e.
\begin{equation}
	\begin{split}
		\amp_{0,\text{fin}}^{(S)}
		&= \amp_{0}^{(S)} \,, \\
		\amp_{1,\text{fin}}^{(S)}
		&= \amp_{1}^{(S)} - \IR_1^{(S)} \amp_{0}^{(S)} \,, \\
		\amp_{2,\text{fin}}^{(S)}
		&= \amp_{2}^{(S)} - \IR_2^{(S)} \amp_{0}^{(S)} - \IR_1^{(S)} \amp_{1}^{(S)} \,.
	\end{split}
	\label{eq:ampFin}
\end{equation}
It is convenient to change loop-momentum integration measure of all bare amplitudes such that all the transcendental coefficients of $\ep^{-n}$ poles of one-loop and two-loop amplitudes are expressed purely in terms of $\log^n(\cdot)$ functions and $\zeta_n$ values, without any remaining dependence on $\log(4\pi)$ and $\gamma_E$.
This can be achieved by changing the overall loop momentum integration measure
\begin{equation}
	\amp_{L,b} = \frac{\ddk_F}{\ddk_P} \tilde{\amp}_{L,b} \,,
	\label{eq:ampTilde}
\end{equation}
where $\ddk_F=\frac{1}{(2\pi)^d}$ is the Feynman measure, and $\ddk_P=\frac{e^{\ep \gamma_E}}{i\pi^{d/2}}$.
Note that
\begin{equation}
	\frac{\ddk_F}{\ddk_P} = f S_\ep \,,
\label{eq:ddk}
\end{equation}
where $S_\ep = (4\pi)^\ep e^{-\gamma_E \ep}$.
After the change of integration measure from $\ddk_F$ to $\ddk_P$, the finite part of the amplitude $\amp_{L,\text{fin}}^{(S)}$ can be related to the bare amplitudes $\amp_{L,b}^{(S)}$ as 
\begin{equation}
	\begin{split}
		\frac{1}{c_0} \tilde{\amp}_{0,\text{fin}}^{(S)}
		&= \tilde{\amp}_{0,b} \,, \\
		\frac{1}{c_1} \tilde{\amp}_{1,\text{fin}}^{(S)}
		&= \tilde{\amp}_{1,b} 
		+ \frac{2 \ci \mu^{-2\ep} \, (q \, \UV_1^{(S)} + \UV_0^{(S)} \, \IR_1^{(S)})}{S_\ep \, (\UV_0^{(S)})^2} \tilde{\amp}_{0,b} \,, \\
		\frac{1}{c_1^2} \tilde{\amp}_{2,\text{fin}}^{(S)}
		&= \tilde{\amp}_{2,b} 
		+ \frac{2 \ci \mu^{-2\ep} ((1+q) \, \UV_1^{(S)} + \UV_0^{(S)} \, \IR_1^{(S)})}{S_\ep \, (\UV_0^{(S)})^2} \tilde{\amp}_{1,b} \\
		&+ \frac{2 \mu^{-4\ep} (q \, (1-q) \, (\UV_1^{(S)})^2 - 2 q \, \UV_0^{(S)} \, \UV_2^{(S)})}{S_\ep^2 \, (\UV_0^{(S)})^4} \tilde{\amp}_{0,b} \\
		&- \frac{2 \mu^{-4\ep} ( 2 q \, \UV_0^{(S)} \, \UV_1^{(S)} \, \IR_1^{(S)} - 2 \, (\UV_0^{(S)})^2 \, \IR_2^{(S)})}{S_\ep^2 \, (\UV_0^{(S)})^4} \tilde{\amp}_{0,b} \,,
	\end{split}
	\label{eq:ampFin2bareTilde}
\end{equation}
where $c_1 = c_0 \frac{\ci}{2} S_\ep \mu^{2\ep} \, \UV_0^{(0)}$.

\section{Pure scheme for the amplitude singularities}
\label{sec:pure}

In this section, we show that the singular part of an amplitude in the introduced \textit{pure scheme} $P$ can be written as a linear combination of amplitude-compatible Feynman integrals with $\ep$-independent coefficients, i.e.
\begin{equation}
    \tilde{\amp}_{L,b} - \frac{1}{c_1^L} \tilde{\amp}_{L,\text{fin}}^{(P)} = \sum_{i} c_{L,i} \, I_{L,i} \,.
\end{equation}
On the one hand, we expect this property to hold because both singular and non-singular terms in the amplitude originate in Feynman diagrams.
On the other hand, the usual derivation of the IR operators $\IR^{(S)}_{L}$ is performed in SCET which provides the integrals $I_{L,i}$ in an amplitude-incompatible representation using linear propagators instead of the standard quadratic ones.
Specifically, we will require for Feynman integrals $I_{L,i}$ in the pure scheme to be subsectors of the diagrammatical topologies appearing in the bare amplitude.
Explicitly, we will focus the bulk of the proof on a massless QCD, but we expect for a similar argument to hold in general.
As a byproduct, we describe an ansatz reconstruction method which will play an important role in the next section.
We will also comment there on the applications of the constructed pure scheme.

Let us denote by $C$ the Catani IR regularization scheme~\cite{Catani:1998bh} defined in the $\overline{\text{MS}}$ UV renormalization scheme, and by $P$ the pure IR regularization and UV renormalization scheme which we construct here.
As a first step towards defining pure UV and IR operators, let us analyse the $S_\ep$ and $\mu$ contributions in Eq.~\eqref{eq:ampFin2bareTilde}.
Since, bare amplitudes $\tilde{\amp}_{L,b}$ computed in $\ddk_P$ measure do not depend on $S_\ep$, the UV $\UV_L^{(S)}$ and IR $\IR_L^{(S)}$ operators have to be defined such that $S_\ep$ cancels out in all $\ep^{-n}$ poles.
The way of canceling the $S_\ep$-dependence is not unique.
In Catani scheme, $S_\ep$ is canceled directly in the prefactors of bare amplitudes $\tilde{\amp}_{L,b}$ in Eq.~\eqref{eq:ampFin2bareTilde}.
Indeed, all UV operators $\UV_{L}^{(C)}$ involve exactly one negative power of $S_\ep$, while the dependence of $\IR_{L}^{(C)}$ on $S_\ep$ cancels.
In pure scheme $P$, we want for all UV $\UV_{L}^{(P)}$ and IR $\IR_{L}^{(P)}$ operators, which we will collectively denote as $\mathcal{O}_{L}^{(P)}$, to arise from $L$-loop Feynman integrals.
Therefore, we choose for each UV or IR operators $\mathcal{O}_{L}^{(P)}$ to involve exactly $L$ positive powers of $S_\ep$.
In summary, we write
\begin{equation}
	\begin{split}
		\UV_{L}^{(C)} &= S_\ep^{-1} \, \bar{\tilde{\UV}}_{L}^{(C)} \,, \\
		\UV_{L}^{(P)} &=  S_\ep^L \, \bar{\tilde{\UV}}_{L}^{(P)} \,, \\
		\IR_{L}^{(C)} &= \bar{\tilde{\IR}}_{L}^{(C)} \,, \\
		\IR_{L}^{(P)} &=  S_\ep^L \, \bar{\tilde{\IR}}_{L}^{(P)} \,.
	\end{split}
	\label{eq:UVIRtilde}
\end{equation}
We introduced here a notation for amplitudes and operators which distinguishes between quantities which have an $\ep$-dependent contribution to its mass dimension.
Since bare amplitudes $\tilde{\amp}_{L,b}$ do not directly depend on $\mu$, we write $\bar{\tilde{\amp}}_{L,b} = \mu^{2L\ep} \tilde{\amp}_{L,b}$, where $\bar{\tilde{\amp}}_{L,b}$ has $\ep$-independent mass dimension, while $\tilde{\amp}_{L,b}$ has $\ep$-dependent mass dimension.
For pure UV and IR operators $\mathcal{O}_{L}^{(P)}$, we demand for $\mu$-dependence to cancel directly in prefactors of bare amplitudes $\tilde{\amp}_{L,b}$ in Eq.~\eqref{eq:ampFin2bareTilde}, i.e. $\bar{\tilde{\mathcal{O}}}_L^{(P)} = \mu^{2L\ep} \tilde{\mathcal{O}}_L^{(P)}$.
Note that since $S_\ep = \frac{\ddk_F}{f \, \ddk_P}$, the pure UV and IR operators $\tilde{\mathcal{O}}_{L}^{(P)}$ are defined in measure $\ddk_P$.

We will now relate pure UV and IR operators $\bar{\tilde{\mathcal{O}}}_{L}^{(P)}$ to the Catani UV and IR operators $\bar{\tilde{\mathcal{O}}}_{L}^{(C)}$.
We start by demanding that the UV poles in both schemes are the same
\begin{equation}
	\bar{\tilde{\UV}}_{L}^{(P)} = \bar{\tilde{\UV}}_{L}^{(C)} + \mathcal{O}(\ep^0) \,.
	\label{eq:UVmatch}
\end{equation}
Then, we derive pure IR operators $\bar{\tilde{\IR}}_{L}^{(P)}$ by using the fact that bare amplitudes $\tilde{\amp}_{L,b}$ in Eq.~\eqref{eq:ampFin2bareTilde} have the same poles in both schemes, i.e.
\begin{equation}
	\small
	\begin{split}
		\bar{\tilde{\IR}}_{1}^{(P)} &= \bar{\tilde{\IR}}_{1}^{(C)} + q (\bar{\tilde{\UV}}_{1}^{(C)}
		- \bar{\tilde{\UV}}_{1}^{(P)}) + \mathcal{O}(\ep^0) \,, \\
		\bar{\tilde{\IR}}_{2}^{(P)} &= \bar{\tilde{\IR}}_{2}^{(C)} - \frac{q}{2} \, (2 \,  \bar{\tilde{\UV}}_{1}^{(C)}  \, \bar{\tilde{\IR}}_{1}^{(C)} + (q-1) \, (\bar{\tilde{\IR}}_{1}^{(C)})^2 + 2 \, \bar{\tilde{\UV}}_{2}^{(C)}) \\
		&+ (\bar{\tilde{\IR}}_{1}^{(C)} + (1+q) \, \bar{\tilde{\UV}}_{1}^{(C)}) \, (\bar{\tilde{\IR}}_{1}^{(P)} + q \, \bar{\tilde{\UV}}_{1}^{(P)}) + q \, \bar{\tilde{\UV}}_{2}^{(P)} \\
		&- (\bar{\tilde{\IR}}_{1}^{(P)})^2 - (1+q) \, \bar{\tilde{\UV}}_{1}^{(P)} \, \bar{\tilde{\IR}}_{1}^{(P)} - \frac{q(q+3)}{2} \, (\bar{\tilde{\UV}}_{1}^{(P)})^2 + \mathcal{O}(\ep^0) \,.
	\end{split}
	\label{eq:IRmatch}
\end{equation}
Note that the pure two-loop IR operator $\bar{\tilde{\IR}}_{2}^{(P)}$ does not depend on the one-loop bare amplitude $\tilde{\amp}_{1,b}$, as Eq.~\eqref{eq:ampFin2bareTilde} may indicate.
Indeed, since the bare one-loop amplitude $\tilde{\amp}_{1,b}$ is multiplied by a linear combination of operators with poles which cancel each other, then the finite part of the bare one-loop amplitude will never contribute to the two-loop IR poles in $\bar{\tilde{\IR}}_{2}^{(P)}$.
Thus, the one-loop bare amplitude $\tilde{\amp}_{1,b}$ contribution to $\tilde{\amp}_{2,\text{fin}}$ in the scheme-matching condition~\eqref{eq:IRmatch} can be substituted with its poles in $\ep$.
We choose these poles to be represented in the pure scheme.

In order to find an explicit representation of the pure UV and IR operators defined in the matching equations~\eqref{eq:UVmatch} and~\eqref{eq:IRmatch}, we use the result for Catani UV and IR operators in massless QCD~\cite{Catani:1998bh,Becher:2009cu,Becher:2009qa}
\begin{equation}
	\small
	\begin{split}
		\bar{\tilde{\UV}}_{0}^{(C)}(\ep) &= 1 \,, \\
		\bar{\tilde{\UV}}_{1}^{(C)}(\ep) &= \frac{\beta_0}{\ep} \,, \\
		\bar{\tilde{\UV}}_{2}^{(C)}(\ep) &= \frac{\beta_0^2}{\ep^2} - \frac{\beta_1}{2\ep} \,, \\
		\bar{\tilde{\IR}}_{1}^{(C)}(\ep)
		&= \frac{e^{\ep\gamma_E}}{2 \, \Gamma(1-\ep)}\,
		\sum_{(i,j)} \, \T_i \circ \T_j \, \left( \frac{1}{\ep^2} 
		- \frac{\gamma_0^i}{2 C_i \, \ep} \right)
		\left( \frac{\mu^2}{-s_{ij}} \right)^\epsilon \,, \\
		\bar{\tilde{\IR}}_{2}^{(C)}(\ep)
		&= - \frac{1}{2} \, \IR_{1}^{(C)}(\ep)
		\left( \IR_{1}^{(C)}(\ep) + \frac{\beta_0}{\ep} \right) \\
		&+\frac{e^{-\ep\gamma_E}\,\Gamma(1-2\ep)}{\Gamma(1-\ep)} 
		\left( \frac{\gamma_1^{\rm cusp}}{8} + \frac{\beta_0}{2\ep} \right) 
		\IR_{1}^{(C)}(2\ep)
		+ H(\ep) \,,
	\end{split}
	\label{eq:}
\end{equation}
with
\begin{equation}
	\footnotesize
	\begin{split}
		H(\ep) 
		&= \frac{1}{16\ep} \, \sum_i \bigg( \gamma_1^i 
		- \frac{1}{4} \, \gamma_1^{\rm cusp} \, \gamma_0^i
		+ \frac{\pi^2}{16}\,\beta_0\,\gamma_0^{\rm cusp}\,C_i \bigg) \\
		&+ \frac{\ci f^{abc}}{24\ep}\,
		\sum_{(i,j,k)} \T_i^a\,\T_j^b\,\T_k^c\,
		\log\frac{-s_{ij}}{-s_{jk}} \log\frac{-s_{jk}}{-s_{ki}} 
		\log\frac{-s_{ki}}{-s_{ij}}  \\
		&- \frac{\ci f^{abc}}{128\ep} \,\gamma_0^\text{cusp}
		\sum_{(i,\,j,\,k)}\T_i^a\,\T_j^b\,\T_k^c\;
		\bigg( \frac{\gamma_0^i}{C_i} - \frac{\gamma_0^j}{C_j} \bigg)
		\log\frac{-s_{ij}}{-s_{jk}}\,\log\frac{-s_{ki}}{-s_{ij}} \,,
	\end{split}
	\label{eq:H}
\end{equation}
where the lower indices $i,j,k$ label coloured external particles in the amplitude $\amp$, $s_{ij} = (p_i+p_j)^2$ are Mandelstam variables, $p_i^\nu$ is the momentum of a particle $i$,  $\sum_{(\vec{i})}$ denotes a sum over all nonequal indices $\vec{i}$, the sum over upper color indices $a,b,c$ is implicit, $f^{abc}$ are the structure constants, $\T^a_i$ is an operator acting on particle $i$ in color space~\cite{Catani:1998bh}, $C_i = \T_i^2$ is the quadratic Casimir in a representation of particle $i$, while the remaining transcendental numbers are the $L$-loop coefficients in perturbative expansion of the beta function $\beta_{L-1}$, the cusp anomalous dimension $\gamma_{L-1}^{\rm cusp}$, and the anomalous dimension $\gamma_{L-1}^{i}$ of particle $i$, and they depend on a scheme for treating external momenta and helicity.

Now, we are going to find the integrand form of the pure UV and IR operators defined in the scheme-matching equations~\eqref{eq:UVmatch} and~\eqref{eq:IRmatch}.
The procedure will be to write an ansatz for the integrand in the form of a linear combination of known integrands with unknown coefficients, and then match the poles to the scheme-matched integrated form.
For the integrand to be pure, we will demand it to be a subtopology of a Feynman-diagrammatic representation of the full amplitude, in order to reproduce all the transcendental functions and values purely from the loop-momentum integration, and thus for the coefficients in the ansatz to be $\ep$-independent purely rational functions of color and kinematics.
Note that our characterization of the pure property is not related to the pure property introduced for integrals of uniform transcendental weight~\cite{Henn:2013pwa}.
For convenience, we are going to consider $\tilde{\mathcal{O}}_{L}^{(P)}$ rather than $\bar{\tilde{\mathcal{O}}}_{L}^{(P)}$, such that all the ansatz integrands have $-2L\ep$ contribution to their mass dimension, just as Feynman diagrams do.
As such, some of the integrands may depend on the scale $\mu$.
We will discuss how to represent these integrals diagrammatically when showing the results at the end of this section.

Let us start by writing the integrated form of the pure UV and IR operators stemming from the scheme-matching equations~\eqref{eq:UVmatch} and~\eqref{eq:IRmatch}
\begin{equation}
	\begin{split}
		\tilde{\UV}_{0}^{(P)} &= 1 \,, \\
		\tilde{\UV}_{1}^{(P)} &= \frac{\beta_0}{\ep} + \mathcal{O}(\ep^0) \,, \\
		\tilde{\UV}_{2}^{(P)} &=  \left( \frac{\beta_0^2}{\ep^2} - \frac{1}{\ep} \left( \frac{\beta_1}{2} + 2 \beta_0^2 \, \log(\mu^2) \right) \right) + \mathcal{O}(\ep^0) \,, \\
		\tilde{\IR}_{1}^{(P)} &= \sum_{(i,j)} \T_i \circ \T_j \, \left( \frac{1}{2\ep^2} - \frac{1}{\ep} \left( \frac{\gamma_0^i}{4 C_i} + \frac{\log(-s_{ij})}{2}\right) \right) + \mathcal{O}(\ep^0) \,, \\
		\tilde{\IR}_{2}^{(P)} 
		&= \,\, w_0^{(0)} 
		+ \sum_{i} w_i^{(1)}
		+ \sum_{(i,j)} \T_i \circ \T_j \, w^{(2)}_{ij} \\
		&+ \sum_{(i,j)} \T_i \circ \T_j \sum_{(k,l)} \T_k \circ \T_l \, w^{(4)}_{ijkl} \\
		&+ \ci f^{abc} \sum_{(i,j,k)} \T^a_i \, \T^b_j \, \T^c_k \, w^{(3)}_{ijk} + \mathcal{O}(\ep^0) \,, \\
	\end{split}
	\label{eq:integrated}
\end{equation}
where $\tilde{\IR}_{2}^{(P)}$ with the explicit form of coefficients $w^{(n)}$ is given in the ancillary files.
Similarly as in the $\ep$-expanded form of $\tilde{\IR}_{2}^{(C)}$, the coefficients $w^{(n)}$ depend on transcendental functions $\log^k(\cdot)$, rational numbers $C_i$, and on transcendental numbers in $\beta_{L-1}$, $\gamma_{L-1}^{\rm cusp}$, $\gamma_{L-1}^{i}$.
We are going to demand that the integrand ansatz integrates to these expressions up to terms finite in $\ep$.

Let us first state the integrand ansatz and then describe its properties
\begin{equation}
	\begin{split}
		\tilde{\UV}_{1}^{(P)} &=  \sum_{n}^{} a_{n} \, I^{(1,2)}_n(-\mu^2) \,, \\
		\tilde{\UV}_{2}^{(P)} &=  \sum_{n}^{} b_{n} \, I^{(2,2)}_n(-\mu^2) \,, \\
		\tilde{\IR}_{1}^{(P)} &= \sum_{(i,j)} \T_i \circ \T_j \sum_{n} d_{ijn} \, I^{(1,3)}_n(s_{ij}) \,,
	\end{split}
	\label{eq:ansatz}
\end{equation}
where $\tilde{\IR}_{1}^{(P)}$ can be schematically depicted as
\\
\centerline{\includegraphics[width=0.2\textwidth]{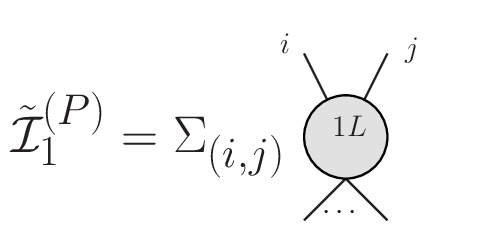}}
\\
while 
\begin{widetext}
\begin{equation}
	\begin{split}
		\tilde{\IR}_{2}^{(P)} 
		=& \,\, c^{(0)} I^{(2,2)}(-\mu^2)
		+ \sum_{i} \sum_{n} c^{(1)}_{in} \, I^{(2,2)}_n(-\mu^2)
		+ \sum_{(i,j)} \T_i \circ \T_j \left( \sum_{n} c^{(2,1)}_{ijn} \, I^{(2,3)}_n(s_{ij}) 
		+ \sum_{n} c^{(2,2)}_{ijn} \, I^{(1,3)}_n(-\mu^2) \, I^{(1,2)}_n(s_{ij}) \right) \\
		&+ \sum_{(i,j)} \T_i \circ \T_j 
		\sum_{(k,l)} \T_k \circ \T_l 
		\sum_{n} c^{(4)}_{ijkln} \, I^{(1,3)}_n(s_{ij}) \, I^{(1,3)}_n(s_{kl})
		+ \ci f^{abc}\sum_{(i,j,k)} \T^a_i \, \T^b_j \, \T^c_k \sum_{n} c^{(3)}_{ijk} \, I^{(2,4)}_n(s_{ij},s_{jk},s_{ki})
	\end{split}
	\label{eq:ansatzIR2}
\end{equation}
can be schematically depicted as
\\
\centerline{\includegraphics[width=0.8\textwidth]{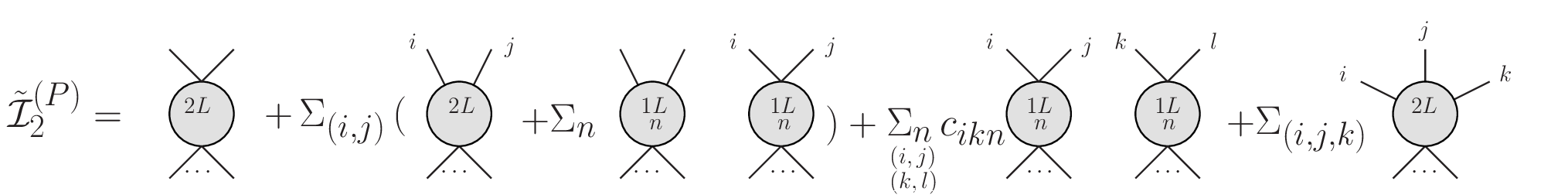}}
\\
Here, each blob diagram consists of a sum of Feynman integrals with $\ep$-independent rational function coefficients.
\end{widetext}
We sum over Feynman a set of integrals $I^{(L,N)}_n$ with $L$ loops and at most $N$ external legs in loop-integration measure $\ddk_P$ which are subsectors of integral topologies relevant for massless parton scattering, as shown in the above diagrammatic representation.
The coefficients $a_{n}, b_{n}, d_{ijn}, c^{(0)}, c^{(1)}_{in}, c^{(2,1)}_{ijn}, c^{(2,2)}_{ijn}, c^{(4)}_{ijkln}, c^{(3)}_{ijk}$ are unknown rational functions.
The choice of the $(L,N)$ pair for each color factor of each pure operator arises from finding the smallest values of $(L,N)$ for which the kinematics matches the corresponding integrated form~\eqref{eq:integrated}.
We also require that each diagram is a subtopology of the whole multi-point amplitude $\amp$.
Alternatively, the $(L,N)$ pair choice can be inferred from the way in which the UV and IR operators are usually computed in the Catani scheme.
Indeed, since UV operators are usually computed from the off-shell two-point function, then we have one-scale $(L,2)$ ansatz for $\tilde{\UV}_{L}^{(P)}$.
Similarly, since IR operators are usually computed from correlators of $L+1$ Wilson lines in SCET~\cite{Becher:2009cu,Becher:2009qa}, then we have at most three-scale $(2,4)$ one off-shell ansatz for $\tilde{\IR}_{2}^{(P)}$, as we consider all internal lines to be massless.
In contrary to SCET propagators, which are linear in loop momenta, we seek a subtopology of the full amplitude $\amp$, i.e. with standard propagators involving squared loop momenta.
Note that the way in which $\log^n(-s_{ij})$ and $\log^n(\mu^2)$ contribute to the integrated form of the pure operators in Eq.~\eqref{eq:integrated} constrains the ansatz for $\tilde{\IR}_{2}^{(P)}$.
For example, $w^{(2)}_{ij}$ requires contributions from both $(-s_{ij})^{-2\ep} c^{(2,1)}_{ijn}$ and $(\mu^2)^{-\ep} (-s_{ij})^{-\ep} c^{(2,2)}_{ijn}$ in order to exponentiate correctly.
In addition, note that $w^{(4)}_{ijkl}$ cannot be expressed in a fully factorized form $(\sum_n I^{(1,3)}_n(s_{ij}) \, c^{(4,1)}_{ijn}) \, (\sum_m I^{(1,3)}_m(s_{kl}) \, c^{(4,2)}_{klm})$ but rather as $\sum_n I^{(1,3)}_n(s_{ij}) \, I^{(1,3)}_n(s_{kl}) \, c^{(4)}_{ijkln}$.
The reason for it is related to the constraint coming from the minimal basis problem~\cite{Chicherin:2020oor,Chicherin:2021dyp}, so let us elaborate on it now.

It is well known, that all Feynman integrals belonging to a fixed integral topology can be related using the Integration-By-Parts (IBP) reduction~\cite{Chetyrkin:1981qh,Tkachov:1981wb,Gehrmann:1999as,Laporta:2000dsw} to a finite set of linearly independent integrals among them, called Master Integrals (MIs)~\cite{Smirnov:2010hn}.
For the integral topologies that we consider here, the resulting MIs can be expanded in $\ep$ into $\sum_{n \geq 0} \ep^{-2L+n} T_{in}$, where $T_{in}$ is a function of at most transcendental weight~\cite{Henn:2013pwa} $n$, and $i$ labels the MIs.
The minimal basis~\cite{Chicherin:2020oor,Chicherin:2021dyp} is defined as a minimal set of combinations of Chen iterated integrals~\cite{Chen:1977oja} which spans all the transcendental functions $T_{in}$ for all MIs at fixed order in $\ep$.
Since the minimal basis depends on the underlying integral topology, some iterated integrals may only appear in this topology in a fixed combination with some other iterated integrals.
On the one hand, the resulting constraint on the number of independent iterated integrals in a topology is very beneficial for numerical evaluation~\cite{Chicherin:2020oor,Chicherin:2021dyp} and amplitude computation~\cite{Gehrmann:2024tds}.
On the other hand, it complicates our ansatz reconstruction procedure.
Indeed, the minimal basis constraints appearing in the MIs are also present in all of the integrals in the corresponding topology by mixing with MI coefficients.
These constraints may be so rigid for a given topology that it cannot span a generic ansatz.
For example, the factorized product of two one-loop topologies, mentioned in the end of the previous paragraph, generates all the expected transcendental terms but their coefficients are too constrained to match the full integrated form $w^{(4)}_{ijkl}$.

Let us now describe the linear reconstruction of unknown coefficients in the integrand ansatz in equations~\eqref{eq:ansatz} and~\eqref{eq:ansatzIR2}.
Note that the known coefficients $w^{(i)}$ in the integrated form~\eqref{eq:integrated} are not of pure weight, i.e. they do not have the same transcendental weight in each internal summand.
Indeed, we have $\gamma_{1}^{\rm cusp} = \gamma_{1,0}^{\rm cusp} + \gamma_{1,2}^{\rm cusp} \pi^2$ and $\gamma_{1}^{i} = \gamma_{1,0}^{i} + \gamma_{1,2}^{i} \pi^2 + \gamma_{1,3}^{i} \zeta_3$, where transcendental numbers $\pi^2$ and $\zeta_3$ have been completely factored out from rational numbers.
Since we do not allow $\ep$-dependence in the ansatz coefficients, then an ansatz consisting only of canonical MIs~\cite{Henn:2013pwa} may not be complete.
In order to ensure the completeness of the ansatz set $I_n^{(L,N)}$, each sum over $n$ should range over all subsectors and crossings of all possible $L$-loop $N$-point diagrammatic topologies.
The ansatz basis may further consist of integrals with Irreducible Scalar Products (ISPs) or dots, i.e. higher propagator powers, up to some overall total rank.
For the pure scheme UV and IR operators, we prefer choosing ISPs over dots because Feynman diagrams of the original amplitude $\amp$ can only have up to $L-1$ dots.

As discussed above, for the ansatz in equations~\eqref{eq:ansatz} and~\eqref{eq:ansatzIR2}, it is sufficient to consider four-point kinematics with one off-shell leg at two loops and three-point kinematics with one off-shell leg at one loop, both with fully massless internal lines.
After performing the IBP reduction for both planar and nonplanar topologies in all crossings with \texttt{Reduze}~\cite{Studerus:2009ye,vonManteuffel:2012np}, we substitute the MIs computed in Refs~\cite{Gehrmann:2000zt,Gehrmann:2001ck,Gehrmann:2023etk}.
A generic form of the integrated ansatz reads
\begin{equation}
	\begin{split}
		&I_n^{(2,4)}(s_{ij},s_{jk},s_{ki}) = (-s_{ijk})^{-2\ep} \sum_{n=0}^{3} \ep^{-4+n} \times \\	
		& \sum_{m, \vec{\alpha}, \vec{\beta} \, : \, m+|\vec{\alpha}|+|\vec{\beta}| \leq n} r_{m \vec{\alpha} \vec{\beta}} \, \zeta_m \, G\left(\vec{\alpha},\frac{s_{ki}}{s_{ijk}}\right) \, G\left(\vec{\beta},\frac{s_{jk}}{s_{ijk}}\right) \,,
	\end{split}
	\label{eq:ansatzGPLs}
\end{equation}
where $s_{ijk} = s_{ij} + s_{jk} + s_{ki}$, $r_{m \vec{\alpha} \vec{\beta}}$ are known rational functions of color and kinematic invariants, $G(\cdot)$ are Generalized Polylogarithms (GPLs)~\cite{Goncharov:1998kja,Gehrmann:2000zt} with letters $\alpha_i \in \{0,1\}$ and $\beta_i \in \{0,1,\frac{s_{ij} + s_{jk}}{s_{ijk}},-\frac{s_{ki}}{s_{ijk}}\}$, while for subsectors with $N \in \{2,3\}$, we have
\begin{equation}
	\small
	\begin{split}
		&I_n^{(L,N)}(s_{ij}) + \mathcal{O}(\ep^0) = (-s_{ij})^{-L\ep} \ep^{L(1-N)} \times \\
		&(r_{n0} + \ep \, r_{n1} + \ep^2 (r_{n20} + r_{n22} \pi^2) + \ep^3 (r_{n30} + r_{n32} \pi^2 + r_{n33} \zeta_3) \,,
	\end{split}
	\label{eq:ansatzZeta}
\end{equation}
where $r_{ij}$ are known rational functions.
We used \texttt{Mathematica} package \texttt{PolyLogTools}~\cite{Duhr:2019tlz} to manipulate GPLs.
Note that in order to match $w^{(3)}_{ijk}$, the unknown coefficients $c^{(3)}_{ijk}$ have to arrange to cancel all of the GPL terms in the integrated ansatz~\eqref{eq:ansatzGPLs} except for simple $\log(\cdot)$ contributions present in the H term~\eqref{eq:H}.
At this point, the ansatz for the pure UV and IR operators, i.e. Eq.~\eqref{eq:ansatz} and Eq.~\eqref{eq:ansatzIR2}, can be directly compared with their known integrated form~\eqref{eq:integrated}.
Since all of the resulting $\ep$-expanded expressions are linear in generic transcendental terms $\ep^{-n} \, \zeta_m \, G\left(\vec{\alpha},\frac{s_{ki}}{s_{ijk}}\right) \, G\left(\vec{\beta},\frac{s_{jk}}{s_{ijk}}\right)$, their corresponding purely-rational coefficients should match.
This creates a large system of linear equations for unknown coefficients $a_{n}, b_{n}, d_{ijn}, c^{(0)}, c^{(1)}_{in}, c^{(2,1)}_{ijn}, c^{(2,2)}_{ijn}, c^{(4)}_{ijkln}, c^{(3)}_{ijk}$.
Let us now turn to their solutions as rational functions of kinematic and color invariants.

As a result of the outlined procedure, we not only restore the well-known Feynman integrand expressions for one-loop UV, two-loop UV, and one-loop IR pure operators
\begin{equation}
	\begin{split}
		\tilde{\UV}_{1}^{(P)} &= \beta_0 \, \includegraphics[width=0.035\textwidth]{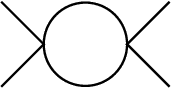}(-\mu^2) \,, \\
		\tilde{\UV}_{2}^{(P)} &= 2(10\beta_0^2+\beta_1) \, \includegraphics[width=0.035\textwidth]{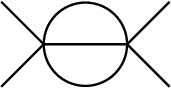}(-\mu^2)
		+ 2\beta_0^2 \, \includegraphics[width=0.035\textwidth]{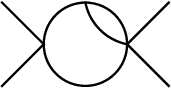}(-\mu^2) \,, \\
		\tilde{\IR}_{1}^{(P)} 
		&= s_{12}^\ep \sum_{(i,j)}^N \T_i \circ \T_j \frac{1}{2} \left( s_{ij} \, \includegraphics[width=0.035\textwidth]{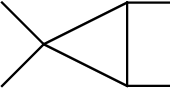}(s_{ij}) 
		+ \frac{\gamma_0^i}{2 C_i} \, \includegraphics[width=0.035\textwidth]{pictures/jaxoDiags/U1.eps}(s_{ij}) \right) \,,
	\end{split}
	\label{eq:reconstructed}
\end{equation}
but we also provide a new expression the two-loop IR pure operator $\tilde{\IR}_{2}^{(P)}$ in the ancillary files.
Let us describe here the properties of the integrand of $\tilde{\IR}_{2}^{(P)}$.
We start by noting that the linear system for unknown rational coefficients $a_{n}, b_{n}, d_{ijn}, c^{(0)}, c^{(1)}_{in}, c^{(2,1)}_{ijn}, c^{(2,2)}_{ijn}, c^{(4)}_{ijkln}, c^{(3)}_{ijk}$ is underdetermined.
Indeed, when considering the full set of ansatz integrals, it leaves some coefficients unconstrained, while, when reducing the ansatz to only some specific topologies or crossings belonging to the respective $(L,N)$ kinematics, it is insufficient.
The property of the integrand solution being not unique is expected since we matched only poles of the integrated form of the ansatz, i.e. any finite Feynman integral can be freely added without spoiling the constraints.
For simplicity, we chose the set of ansatz integrals such that $w^{(3)}$ in the ancillary files contains only planar integrals and all the unconstrained integral coefficients $c$ are set to 0 to reduce the number of contributing integrals.
In particular, the sum in $w^{(0)}, w^{(1)}, w^{(2)}, w^{(3)}$ runs through 1, 6, 7, 17 two-loop integrals, with additional 2 and 4 products of two one-loop integrals in $w^{(2)}$ and $w^{(4)}$, respectively.
Of course, it is always possible to consider a larger set of ansatz integrals, e.g.  to include also the nonplanar contributions to $w^{(3)}$, however, it might result in generating rational functions of high degree and with large fraction coefficients.
Contrarily, considering an even smaller set of ansatz integrals leads to an incomplete system.
This includes requiring e.g. only one specific crossing contribution to $w^{(3)}$, or only planar contributions to $w^{(0)}$, $w^{(1)}$ or $w^{(2)}$, or a fully-factorized form of $w^{(4)}_{ijkl} = (\sum_n I^{(1,3)}_n(s_{ij}) \, c^{(4,1)}_{ijn}) \, (\sum_m I^{(1,3)}_m(s_{kl}) \, c^{(4,2)}_{klm})$, as mentioned earlier.
The last two are equivalent to the statement that neither two-loop planar nor a product of two sums of one-loop three-point internally massless integrals with one leg off-shell can span all the $\ep^{-n} \zeta_m$ terms below weight 4 with arbitrary rational coefficients.
It is due to too rigid minimal basis constraints. 

Let us also elaborate on further properties of the reconstructed one-loop and two-loop pure UV and IR operators.
Firstly, in contrary to the two-loop IR operator in the Becher-Neubert scheme~\cite{Becher:2009cu,Becher:2009qa}, $\tilde{\IR}_{2}^{(P)}$ has a nontrivial contribution $w^{(3)}$ from the color tripole, as in the Catani scheme~\cite{Catani:1998bh}.
Since there is a nonzero contribution from the color quadrupole to the three-loop IR operator in the Becher-Neubert scheme~\cite{Almelid:2015jia}, we expect for higher ($L+1$)-fold multipole terms to appear in $\tilde{\IR}_{L}^{(P)}$.
Secondly, we note that there is a minor dependence of a purely-UV origin on $q$ in $w^{(0)} = -\frac{4 q \beta_0^2}{\ep}$, which is absent in both Catani and Becher-Neubert schemes.
Thirdly, as mentioned earlier, it is possible to find a representation of the pure operators as subdiagrams of the full amplitude $\amp$, despite the dependence on $\mu^2$.
Indeed, it is enough to consider $\left(\frac{s}{\mu^2}\right)^{-2L\ep} \tilde{\mathcal{O}}_{L}^{(P)}$ instead of $\tilde{\mathcal{O}}_{L}^{(P)}$, where $s$ is some chosen squared scale which has an interpretation corresponding to one of the diagrammatic channels, e.g. a Mandelstam variable.
In the ancillary files, we chose to provide the integral topology definitions parametrized with $s=s_{12}$.

Note that the H term~\eqref{eq:H} simplifies for processes which involve a low number of external legs.
Indeed, below four massless colored external particles, the tripole contribution $w^{(3)}$ vanishes, while for a scattering of exactly four colored particles, without any colorless legs, the kinematics of $\log(\cdot)$ functions simplifies due to an overall momentum conservation $s_{12}+s_{23}+s_{31}=0$.
Therefore, the integrand ansatz for the two-loop pure IR operator $\tilde{\IR}_{2}^{(P)}$ should consists of two-loop four-point integrals with all on-shell legs.
We constructed a separate integrand ansatz for $w^{(3)}$ and we found its integrand form using IBP relations generated with \texttt{Reduze}~\cite{Studerus:2009ye,vonManteuffel:2012np} and MIs computed in Refs~\cite{Smirnov:1999gc,Henn:2013pwa,Bargiela:2021wuy}.
The function space in the $\ep$-expanded ansatz simplifies such that a generic transcendental term reads $\ep^{-n} \, \zeta_m \, G\left(\vec{\alpha},\frac{s_{ij}+s_{jk}}{s_{ij}}\right)$, where $G(\cdot)$ are Harmonic Polylogarithms (HPLs)~\cite{Remiddi:1999ew} with letters $\alpha_i \in \{0,1\}$.
After solving a linear system of conditions for the unknown rational coefficients $c^{(3)}_{ijk}$, we chose a simple representation which consist of 12 integrals which are, similar to the one off-shell case, all planar and they do not contain either dots or ISPs.
We provide this new result for the tripole contribution $w^{(3)}$ in the ancillary files.

Finally, we conclude with two more general comments on the outlined ansatz reconstruction method.
Firstly, it does not restrict to some specific values of the rational numbers $\beta_0$, $\beta_1$, $\gamma_{0}^{\rm cusp}$, $\gamma_{1,0}^{\rm cusp}$, $\gamma_{1,2}^{\rm cusp}$, $\gamma_{0}^{i}$, $\gamma_{1,0}^{i}$, $\gamma_{1,2}^{i}$, $\gamma_{1,3}^{i}$, or the color invariants.
Therefore, a similar construction should also work in different gauge theories, and for other schemes for treating the external particle momenta and helicity e.g. Conventional Dimensional Regularization (CDR), `t Hooft-Veltman (tHV), and the four-dimensional helicity scheme (4DH).
Secondly, the ansatz reconstruction is even simpler when applying to a fixed process.
Indeed, in such case, one can explicitly perform the color operator $\T$ algebra in Eq.~\eqref{eq:integrated}, thus the remaining integrated expressions have fixed kinematic labels $1,2,3,4,5,...$, in contrary to generic summed-over indices $i,j,k,l$.
Then, one can exploit the freedom stemming from the underdetermined system of ansatz-matching equations and choose an integrand representation suited to a particular process.
As an example, we computed the pure UV and IR operators for the $q\bar{q} \to gg$ scattering at two loops in massless QCD.
We used \texttt{qgraf}~\cite{Nogueira:1991ex} to generate Feynman diagrams, \texttt{FORM}~\cite{Vermaseren:2000nd} to perform color and spinor algebra, and the tHV projector decomposition introduced recently in Ref.~\cite{Peraro:2020sfm}.
We found that all poles of the bare two-loop amplitude $\tilde{\amp}_{2,b}$~\cite{Bern:2003ck} as well as of the pure IR operator $\tilde{\IR}_{2}^{(P)}$ can be spanned by the same 27 integrals with at most 1 ISP, including 19 planar and 8 nonplanar integrals, which are the same in both tHV and 4DH scheme.

\section{Locally finite part of an example helicity amplitude}
\label{sec:local}

In this section we show how to turn the finite part of an amplitude defined in Eq.~\eqref{eq:ampFin2bareTilde} into a locally finite expression, i.e.
\begin{equation}
    \frac{1}{c_1^L} \tilde{\amp}_{L,\text{fin}}^{(P)} = \sum_i c_{L,i,\text{fin}} \tilde{M}_{L,i,\text{fin}} \,,
\end{equation}
on an example of the $q\bar{q} \to gg$ scattering.

Let us start by explaining the difference between locally and \textit{globally finite} expressions.
We refer to an expression as globally finite if it sums to a finite quantity while its summands may separately be singular.
It means that the summands have to be computed using a regulator, e.g. $\ep$, and the physical limit $\ep \to 0$ can only be taken at the end of the calculation.
For example, the right-hand side of Eq.~\eqref{eq:ampFin2bareTilde} is globally finite.
Contrarily, we refer to an expression as \textit{locally finite} if each of its summands is separately finite.
In dimReg, it means that each summand can be evaluated explicitly at $\ep=0$ without producing any singularities, even numerically.
For example a massless box Feynman integral is locally finite in $d=6$.

We turn the finite part of an amplitude defined in Eq.~\eqref{eq:ampFin2bareTilde} into a locally finite expression in two steps.
Firstly, we follow Sec.~\eqref{sec:pure} to write the predicted singular part of an amplitude as a sum of amplitude-compatible Feynman integrals with $\ep$-independent coefficients.
Secondly, we subtract these singular terms from the bare amplitude $\tilde{\amp}_{L,b}$, and then we perform an overall IBP reduction onto a common MI basis, i.e.
\begin{equation}
    \frac{1}{c_1^L} \tilde{\amp}_{L,\text{fin}}^{(P)} 
    = \tilde{\amp}_{L,b} + \sum_{j} c_{L,j} \, I_{L,j} 
    = \sum_i c_{L,i} \tilde{M}_{L,i} \,.
\end{equation}
We will now show on an example how to choose the MI basis $\tilde{M}_{L,i}$ such that each MI is locally finite, and how their coefficients $c_{L,i}$ transform.

Consider the leading color contribution to the $(T^{a_3}T^{a_4})_{i_1\bar{i}_2}$ color factor of the $q(p_1) \, \bar{q}(p_2) \to g(-p_3) \, g(-p_4)$ scattering amplitude in two-loop massless QCD.
All the Feynman diagrams of this contribution are planar and have natural leg ordering, together with the crossing $(p_1 \to p_2 \,, p_2 \to p_3 \,, p_3 \to p_4 \,, p_4 \to p_1)$, which we will abbreviate as x1234.
As mentioned above, in the first step, following our ansatz-matching construction in Sec.~\eqref{sec:pure}, we write the singular part of our amplitude as a sum of 15 Feynman integrals, which all belong integral topologies appearing in the as the bare i.e. the double-box and its x1234 crossing.
We attach the resulting singular part of our amplitude in the ancillary files.
Note that the poles of the constructed singular part are the same in any scheme because they have to cancel bare amplitude poles.
While the finite part differs in different schemes, the relation between schemes can be reconstructed from their definition.
In the second step, we IBP reduce both the singular part and the bare amplitude onto a common basis of 13 MIs.
We can choose each MI in this basis to be locally finite
\begin{equation}
\footnotesize
\begin{split}
    \tilde{M}_{2,i,\text{fin}} &\in \{ 
    F_1 \,, 
    F_2 \,, 
    k_2 \cdot p_1 \, F_2 \,,
    F_3 \,,
    k_1 \cdot p_1 \, F_3 \,,
    k_2 \cdot p_1 \, F_3 \,,
    F_4 \,,
    F_5
    \} \\
    &\cup \,
    \{ 
    F_1 \,, 
    F_2 \,, 
    k_2 \cdot p_1 \, F_2 \,,
    F_3 \,,
    k_1 \cdot p_1 \, F_3
    \}|_{x1234} \,,
\end{split}
\end{equation}
where all the scalar products are understood to appear in the numerator of the integrand under the two-loop integral sign.
In the same notation, the locally finite $F_i$ integrals have been derived recently in Ref.~\cite{Gambuti:2023eqh}
\begin{equation}
\scriptsize
\begin{split}
    \vec{F} =& \Biggl\{ 
    \gram {k_2 \, p_1 \, p_2 \choose k_1 \, p_3 \, p_4} \,,\,\,
    \gram {k_1 \, p_1 \, p_2 \, p_3 \choose k_2 \, p_1 \, p_2 \, p_3} \,,\,\,
    \gram (k_1 \, p_3 \, p_4) \, \gram {k_2 \, p_1 \, p_2 \choose p_1 \, p_2 \, p_4} \,,\,\, \\
    &\gram (k_1 \, p_3 \, p_4) \, (k_2-p_1)^2 \,,\,\,
    (k_2-p_1)^2 \, (k_1+p_4)^2
    \Biggr\} \, 
    \times
    \includegraphics[width=0.1\textwidth]{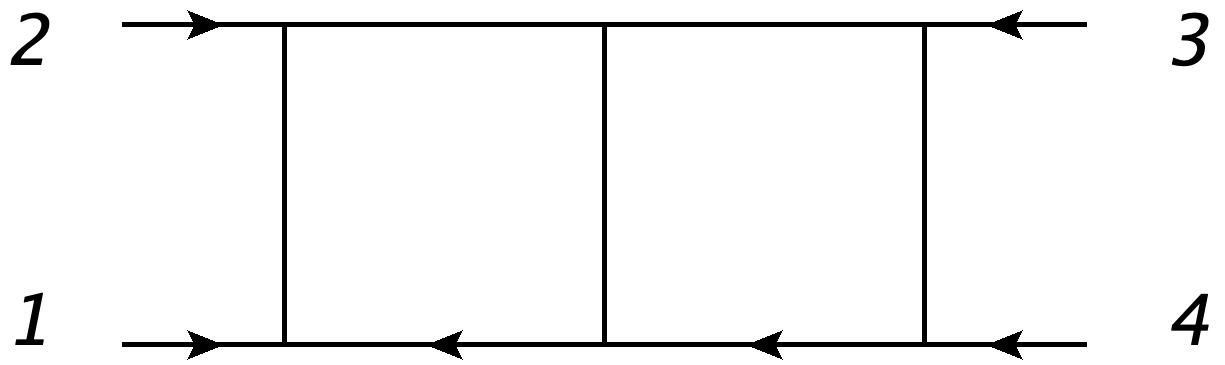} \,,
\end{split}
\end{equation}
with the loop momenta $k_2$ and $k_1$ in the double-box diagram assigned respectively along the internal arrows, the Gram determinants
\begin{equation}
\gram (q_1 \, q_2 \, q_3) = 
\left| \begin{array}{ccc}
q_1 \cdot q_1 & q_1 \cdot q_2 & q_1 \cdot q_3 \\
q_2 \cdot q_1 & q_2 \cdot q_2 & q_2 \cdot q_3 \\
q_3 \cdot q_1 & q_3 \cdot q_2 & q_3 \cdot q_3 
\end{array}
\right|
\end{equation}
and
\begin{equation}
\gram {q_1 \, \dots \, q_n \choose r_1 \, \dots \, r_n} = 
\left| \begin{array}{cccc}
q_1 \cdot r_1  & q_1 \cdot r_2 & \cdots & q_1 \cdot r_n \\
q_2 \cdot r_1  & q_2 \cdot r_2 & \cdots & q_2 \cdot r_n \\
\vdots & \vdots &  \ddots & \vdots \\ 
q_n \cdot r_1  & q_n \cdot r_2 & \cdots & q_n \cdot r_n
\end{array}
\right| \,.
\end{equation}
In this MI basis, for gluon same-helicity configurations, the MI coefficients $c_{L,i,\text{fin}}$ are finite when $\ep \to 0$.
Therefore, we wrote the resulting finite part of helicity amplitude in a locally finite manner, i.e. such that one can put $d=4$ without the $\ep$ regulator and perform the integration numerically.
In comparison, for gluon opposite-helicity configurations, the MI coefficients $c_{2,i}$ are $\mathcal{O}(\ep^{-1})$.
This means that we have to expand our locally finite MIs to $\mathcal{O}(\ep^{1})$ in order to reproduce the full finite part of the corresponding helicity amplitude.
Even though the resulting expression is not locally finite, it is still an improvement in comparison to using this MI basis for the bare amplitude alone.
Indeed, analytically, the MI coefficients are much less singular, i.e. $\mathcal{O}(\ep^{-1})$, for the finite part of the amplitude than for the bare amplitude, which diverge as most as $\mathcal{O}(\ep^{-4})$, which would require expanding the corresponding locally finite MIs to $\mathcal{O}(\ep^{4})$.
In addition, the locally finite MIs are also locally finite order by order in $\ep^n$, which improves the stability of their numerical evaluation.
As a result, one can refer to the resulting MI basis for the gluon opposite-helicity configurations as quasi-finite.

Let us comment on some possible extensions of this analysis.
Firstly, it would be useful to find some general guidance on how to choose both the integrals forming the subtracted singular part of the amplitude and the basis of finite MIs such that their coefficients for the finite amplitude part are always finite.
So far, the existence of MI basis has been proven only for each of these two properties separately, i.e. the one having all locally finite MIs in Refs~\cite{vonManteuffel:2014qoa,vonManteuffel:2015gxa}, or the one having all finite MI coefficients in Ref.~\cite{Chetyrkin:2006dh}.
For example, we chose both these integral types to be subsectors of the double-box topology and its crossing.
Note that there is also beetle~\cite{Gambuti:2023eqh} topology contribution to the Feynman diagrams.
By testing a few example choices of integrals from mixed double-box and beetle topologies, we did not find any beneficial choice.
Alternatively, one could imagine exploiting only those IBP identities which do not generate $\ep$-dependence in the MI coefficient.
These are the identities generated by the external vectors $p_i^\mu$ and not by loop vectors $k_l^\mu$.
Since Feynman rules do not generate any $\ep^{-n}$ poles in the coefficients of the unreduced integrals, then the amplitude after such an altered IBP reduction would also have finite coefficients of the altered integral basis.
Note that, in an incomplete IBP system, the size of the altered integral basis is much larger than of the standard MI set resulting from a complete system.
Moreover, is such case, the dots are not always reducible in terms of ISPs.
Therefore, it is not obvious how to guarantee the local finiteness of the chosen altered integral basis.
Secondly, it might be beneficial to investigate if there is a choice of both the integrals forming the singular part of the amplitude and the basis of finite MIs such that resulting finite part of the virtual contribution is suppressed.
This would mean that the real radiation corrections dominate, and one could approximate the result by neglecting the virtual correction.
If the resulting finite amplitude part was constructed to be vanishing as $\ep \to 0$, then the result in this approximation would become exact.

Finally, it would be interesting to explore if all of the steps necessary for writing the finite part of the amplitude in a locally finite manner can be performed using purely integrand-level identities.
In practice, this would require avoiding the step of reconstructing an integrand ansatz for the singular part of the amplitude from its integrated form.
Instead, one may map the products between the UV/IR operators and the lower-order amplitudes appearing in Eq.~\eqref{eq:ampFin2bareTilde} onto two-loop integral topologies compatible with a given process.
The most challenging problem in this approach would be to avoid introducing additional MIs, resulting from products of two one-loop expressions, in a way that the locally finite MIs with finite coefficients can still be found.
In the considered $q\bar{q} \to gg$ example, these products of one-loop expressions lead to the so called graphless~\cite{vonManteuffel:2012np} integral sectors, e.g. box multiplied with a bubble, which are not simply one-loop-factorizable diagrams for four-point kinematics.

\section{Conclusions}
\label{sec:concl}

In this work is we expressed the singular part of the amplitude in terms of Feynman integrals compatible with topologies appearing in the bare amplitude as a first step towards constructing a locally finite part of the amplitude.
To this end, we proposed a new pure UV renormalization and IR regularization scheme defined such that the UV and IR operators have a form consistent with the Feynman integrand of bare two-loop QCD amplitude.
Note that these pure operators do not have a uniform transcendental weight.
In order to find such an integrand form of the pure UV and IR operators, we exploited an ansatz reconstruction method.
On the example of two-loop massless QCD, we matched their known integrated form with an ansatz of a linear combination of Feynman integrals with unknown rational function coefficients, such that all the transcendental factors i.e. $\ep^{-n}$, $\zeta_m$, and $\log^k(\cdot)$ are generated from Feynman integrals.
As a result, we found a class of pure UV and IR operators which are defined up to an arbitrary finite part in $\ep$.
Due to this nonuniqueness, the exact form of these operators can be suited to a particular process.
Consistently with the SCET argument~\cite{Becher:2009cu,Becher:2009qa}, the pure two-loop IR operator has at most a four-point one off-shell integral topology for any multi-point scattering amplitude.
Beyond two-loop massless QCD, we expect our construction to work in other theories and orders, as long as the singular part of the amplitude can be predicted.

In addition, we chose a basis of locally finite Master Integrals, thus aiming for a locally finite expression for the finite part of the amplitude, on an example of the $q\bar{q} \to gg$ scattering.
We subtracted the predicted singularities from the bare two-loop amplitude in their pure amplitude-compatible form.
Then, we IBP reduced the resulting finite part of the amplitude onto a basis of locally finite MIs.
For some helicity configurations, the MI coefficients are also finite in $\ep$, therefore the whole expression becames locally finite, i.e. one can explicitly put $\ep=0$ at the integrand level.
For other helicity configurations, the MI coefficients are singular at most as $\mathcal{O}(\ep^{-1})$.
In comparison, the MI coefficients are singular up  $\mathcal{O}(\ep^{-4})$ for the bare amplitude in the same MI basis.

There is a plethora of possible future directions which could build on this work.
They would have a substantial impact on understanding the structure of scattering amplitudes both formally and phenomenologically.
Let us elaborate on some of them.

Regarding the pure scheme, firstly, it would be useful to generalize our result to higher-loop orders.
Motivated by the SCET argument~\cite{Gardi:2009qi,Becher:2009qa,Almelid:2015jia}, we expect at most $(L+2)$-point integrals to contribute to the pure $L$-loop IR operator.
Secondly, one could extend our pure scheme to include also massive colored particles.
The integrated expressions for such IR operators are currently available at three loops for the former~\cite{Gardi:2009qi,Becher:2009qa,Almelid:2015jia} and at two loops for the latter~\cite{Catani:2000ef,Becher:2009kw}.
Thirdly, it would be interesting to examine how does the minimal basis put constraints also on the finite part of the amplitude.
Indeed, the existence of the resulting linear relations between rational coefficients of transcendental functions contributing to the physical part of the amplitude would reduce the complexity of amplitude computations.
Fourthly, the pure scheme is defined such that the dependence on the renormalization scale $\mu$ factorizes from the finite part of the amplitude.
This may simplify the estimation of missing higher-order uncertainties (MHOUs) in the perturbative expansion.
Finally, we conclude by pointing out that the outlined method of reconstructing an integrand ansatz from an integrated expression could be applied beyond UV and IR operators e.g. to the finite part of the amplitude or to individual Feynman integrals.

Regarding the locally finite amplitude part, firstly, it would be useful to find some general guidance on how to choose both the integrals forming the singular part of the amplitude and the basis of finite MIs such that their coefficients for the finite amplitude part are always finite.
Secondly, it might be beneficial to investigate if there is a choice of both the integrals forming the singular part of the amplitude and the basis of finite MIs such that the resulting finite part of the virtual correction is suppressed in comparison to the real radiation contribution.
Finally, it would be interesting to explore if all of the steps necessary for writing the finite part of the amplitude in a locally finite manner can be performed using purely integrand-level identities.
This would avoid the step of reconstructing an integrand ansatz for the singular part of the amplitude from its integrated form.

\section*{Acknowledgements}

This work was inspired by discussions with F. Caola on the integrand form of infrared singularities of $q\bar{q} \to gg$ scattering in one-loop massless QCD during the early stages of DPhil studies of the author.
We are grateful to S. Zoia for valuable comments on this draft.
We thank G. Falcioni, J. Haag, P. Jakub\v{c}\'{i}k, M. Marcoli, V. Sotnikov, and T.-Z. Yang for interesting discussions.
This research was supported by the Swiss National Science Foundation (SNF) under contract 200020-204200 and by the European Research Council (ERC) under the European Union's Horizon 2020 research and innovation programme grant agreement 101019620 (ERC Advanced Grant TOPUP).
All Feynman graphs were drawn with \texttt{JaxoDraw}~\cite{Vermaseren:1994je,Binosi:2003yf}.

\bibliographystyle{apsrev4-1}
\bibliography{references}

\end{document}